\documentclass{article}

\usepackage{arxiv}

\usepackage[utf8]{inputenc} 
\usepackage[T1]{fontenc}    
\usepackage{hyperref}       
\usepackage{url}            
\usepackage{booktabs}       
\usepackage{amsfonts}       
\usepackage{nicefrac}       
\usepackage{microtype}      
\usepackage{lipsum}
\usepackage{setspace}
\usepackage{lineno}
\usepackage{amsmath}
\usepackage{graphicx}
\usepackage{float}
\usepackage{xcolor}
\usepackage{lineno}
\usepackage{graphicx}
\usepackage{dcolumn}
\usepackage{bm}
\usepackage{float}
\begin{document}

\title{Dengue Seasonality and Non-Monotonic Response to Moisture: A Model-Data Analysis of Sri Lanka Incidence from 2011 to 2016}

\author{
  Milad Hooshyar\thanks{Corresponding author} \\
  Princeton University\\
  \texttt{hooshyar@princeton.edu}\\
   \And
  Caroline E. Wagner \\
  McGill University \\
   \And
  Rachel E. Baker \\
  Princeton University\\
  \And
  Wenchang Yang \\
  Princeton University\\
  \And
  Gabriel A. Vecchi \\
  Princeton University\\
  \And
  C. Jessica E. Metcalf \\
  Princeton University\\
   \And
  Bryan T. Grenfell \\
  Princeton University \\
   \And
 Amilcare Porporato \\
  Princeton University\\
  }
\date{\today}

\maketitle

\begin{abstract}
Dengue fever impacts populations across the tropics. Dengue is caused by a mosquito transmitted flavivirus and its burden is projected to increase under future climate and development scenarios. The transmission process of dengue virus is strongly moderated by hydro-climatic conditions that impact the vector's life cycle and behavior. Here, we study the impact of rainfall seasonality and moisture availability on the monthly distribution of reported dengue cases in Sri Lanka. Through cluster analysis, we find an association between seasonal peaks of rainfall and dengue incidence with a two-month lag. We show that a hydrologically driven epidemiological model (HYSIR), which takes into account hydrologic memory in addition to the nonlinear dynamics of the transmission process, captures the two-month lag between rainfall and dengue cases seasonal peaks. Our analysis reveals a non-monotonic dependence of dengue cases on moisture, whereby an increase of cases with increasing moisture is followed by a reduction for very high levels of water availability. Improvement in prediction of the seasonal peaks in dengue incidence results from a seasonally varying dependence of transmission rate on water availability.
\end{abstract}

\section{Introduction}
Dengue fever was first mentioned in a Chinese encyclopedia of disease symptoms and remedies published sometime during 265-420 A.D. \cite{gubler1998dengue}. The first recorded dengue epidemic occurred during 1779 and 1780 in Asia, Africa, and North America \cite{rush1951account, hirsch1883dengue, gubler1998dengue}, and its burden has increased dramatically in recent decades, making it a major problem in the tropics, comparable to that of malaria. According to the World Health Organization (WHO) \cite{who}, the number of reported dengue cases has surged from about  500k cases in 2000 to over 4.2 million in 2019 while the death toll has increased from almost 1000 to more than 4000 in the same period. Because dengue is often asymptomatic or with mild symptoms, it is thought to be under-reported, with the actual cases potentially order of magnitudes larger ($~ 390$ million in 2010, as suggested in \cite{bhatt2013global}). There are four distinct strains of dengue virus (DENV), whereby infection to each strain creates life-long immunity only to that particular strain and makes the secondary infections by a new strain very severe \cite{jeewandara2015change, reich2013interactions}. 

Hydro-climatic conditions can greatly impact several aspects of infection transmission and partly control the temporal and spatial patterns of disease spread \cite{rogers2015dengue,bhatt2013global, pascual2000cholera, deyle2016global, wagner2020climatological, mordecai2017detecting, mordecai2019thermal}. The survival probability and transmission of disease agents (e.g., virus, bacterial, and other parasites) are often controlled by hydro-climatic variables such as temperature, rainfall, air humidity \cite{lowen2007influenza}.  In vector-borne diseases, the abundance of breeding sites, larva development, feeding, and survival chance are all affected by climate and water availability \cite{rogers2006climate,yang2009assessing, yamana2013projected, gharbi2011time, lu2009time, wagner2020climatological}. For instance, it is well-known that water availability is a key factor in the transmission cycle of dengue as the vector ( \emph{Aedes aegypti} mosquito) breeds in natural or artificial water containers \cite{hales2002potential, focks1993dynamic, patz1998dengue}. This hydro-climatic dependence highlights the possible increase in the future burden of dengue under climate change. It is estimated that almost $50$–$60\%$  of the (projected) global population in 2085 would be at risk of dengue transmission compared to only $30\%$ in 1990 via the combined effects of population growth and climate change \cite{hales2002potential}.  These predictions also highlight the potential increase in the spread of diseases such as chikungunya, Zika fever, Rift valley fever, and yellow fever for which \emph{Aedes aegypti} mosquito is a common vector \cite{leta2018global}. 

Dengue fever has been circulating in Sri Lanka for the last 40 years \cite{kanakaratne2009severe, wagner2020climatological}. In a recent epidemic in 2017 $186$k confirmed cases were reported with a death toll of almost $~230$, according to Sri Lanka's Health Ministry \cite{epid, wagner2020climatological}. The disease exists in nearly all geographic regions across the country; however, the severity and its seasonal variation are highly spatially variable. Wagner et al.  \cite{wagner2020climatological} studied the association of dengue incidence with rainfall and temperature and found a positive feedback from these climatic variables on the dengue transmission rate with a time lag of several weeks. This analysis allowed for a quantitative evaluation of the future risk of dengue across the island. 

Here, we refine the analysis of Wagner et al.  \cite{wagner2020climatological} analysis to allow mechanistically for the impact of hydro-climatic forcing. Specifically, we use a minimalist coupled hydrologic-epidemiological model (HYSIR) to explore the association between rainfall seasonality patterns (RSPs) and dengue seasonality patterns (DSPs) and the phase difference between the peaks of rainfall and dengue cases. In this model, water availability is modeled by a minimalist bucket model \cite{porporato2004soil}, which is fed by rainfall and maintains a memory of rainfall forcing. The hydrologic and epidemiological models are coupled by assuming a functional dependence between water availability and transmission rate. We show that the introduction of hydrologic memory, in addition to the nonlinear dynamics of the disease model, can explain the observed seasonality of dengue incidence without the addition of a lag parameter as was done in \cite{wagner2020climatological}. Accurate prediction of seasonal peaks of dengue cases can be achieved by assuming a seasonally varying response of transmission rate on water availability. Our results further suggest a non-monotonic relationship between water availability and dengue transmission rate. Such dependence predicts an initial increase in transmission rate followed by a drop under very wet conditions. 

\section{Rainfall seasonality in Sri Lanka}

There are four distinct monsoon seasons that contribute to precipitation in Sri Lanka \cite{thambyahpillay1954rainfall, suppiah1996spatial}. The First Inter-Monsoon (FIM) spans from March through April with rainfall attributed to the disturbances within the Inter-Tropical Convergence Zone (ITCZ). This is followed by the Southwest Monsoon (SWM) which typically begins late in May and lasts until September. The second Inter-Monsoon (SIM) includes rainfall contributions from the tropical depressions or cyclones in the Bay of Bengal and is active during October and November. Lastly, there is the Northeast Monsoon (NEM) from December to February \cite{suppiah1989relationships, suppiah1996spatial}. 

\begin{figure}[h]
    \centering
    \includegraphics[width=1\textwidth]{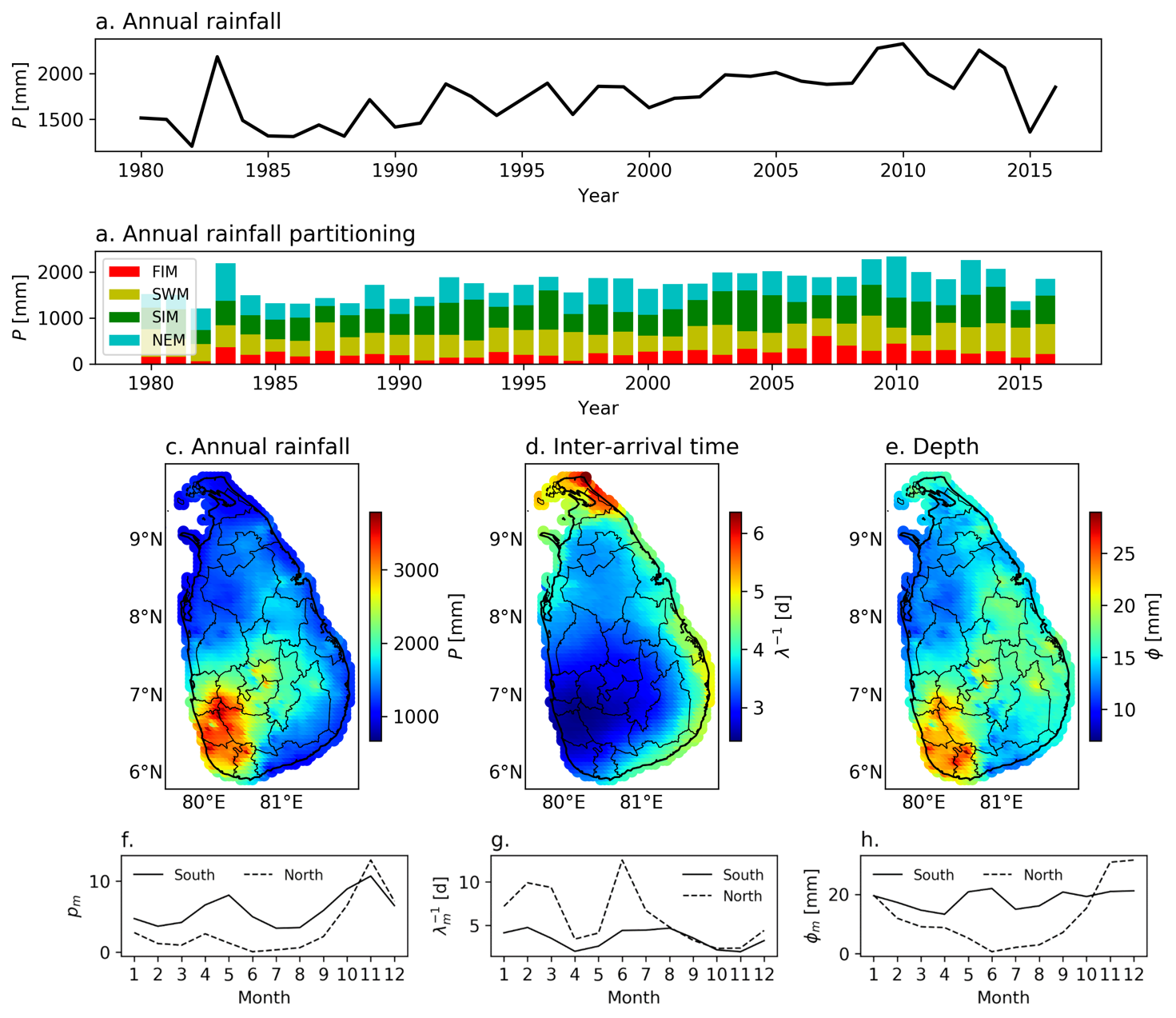}
    \caption{Annual rainfall in Sri Lanka from 1980-2016. (a) and (b) show the spatial average of annual rainfall across the country and the partitioning of rainfall into the monsoons FIM (March and April), SWM (May to September), SIM (October and November), and NEM (December to February).(c), (d), and (e) show the spatial distribution of mean annual rainfall, the inter-arrival time, and depth of rainfall events. The monthly variation of mean daily rainfall, the inter-arrival time, and depth of rainfall events for southern (below the latitude $7$) and northern (above the latitude $9$) regions are shown in (f), (g). and (h). }
    \label{fig:MAR}
\end{figure}

Fig \ref{fig:MAR}a shows the spatial average of annual precipitation across Sri Lanka from 1980-2016. The partitioning of rainfall between the four monsoon seasons (i.e., FIM, SWM, SIM, and NEM) is depicted in Fig. \ref{fig:MAR}b. A slight increasing trend ($~ +27 \text{mm}/\text{y}$) in rainfall exists from 1985-2010 followed by a sharp decrease in rainfall after 2010 ($~ -88 \text{mm}/\text{y}$). Regression analysis reveals that although all monsoon seasons had an increasing rainfall trend in this period, the majority of the overall increase ($~ 68\%$) is attributed to the higher rainfall from FIM and NEM (December to April). This trend is followed by a sharp drop after 2010. Despite this overall decreasing trend, the SWM season had increases in rainfall ($~ +55 \text{mm}/\text{y}$), which were surpassed by a significant drop in the rain during FIM and NEM. This analysis reveals that the inter-annual rainfall trend from 1985-2016 is dominated by the change in rainfall from FIM and NEM (increase in 1985-2010 and decrease in 2010-2016).

The spatial distribution of mean annual rainfall from 1980 to 2016 across the island is shown in Fig.\ref{fig:MAR}c. The annual rainfall ranges from more than $3 \text{m}$ in the southwest to less than $1 \text{m}$ in northern regions. The spatial distribution of rainfall intermittency in terms of inter-arrival time $\lambda^{-1}$ (the expected time in days between rainfall events) and the depth $\phi$ (the expected amount of rain from each event) is shown in \ref{fig:MAR}d and e. The southwestern region has the highest annual rainfall with more frequent and intense rain events. Total rainfall decreases moving toward the north and rainfall events become less frequent (on average every $~6$ days in the northern regions). It is interesting to note that despite relatively comparable average rain depth in northern and eastern regions, the mean annual rainfall in the north is much lower mainly due to less frequent rainfall events.

The seasonality of rainfall from 1980-2009 can be studied quantitatively by clustering the seasonality patterns. To do so, $p_m$ at each location is defined as the ratio of total rain in month $m$ to the total rain during the study period. This gives a vector $p$ with 12 elements summing up to unity which describes normalized seasonal rainfall patterns. These vectors are then fed into a  K-means clustering algorithm to detect the most distinct rainfall seasonality patterns (RSPs) in the island. 

\begin{figure}[h]
    \centering
    \includegraphics[width=1\textwidth]{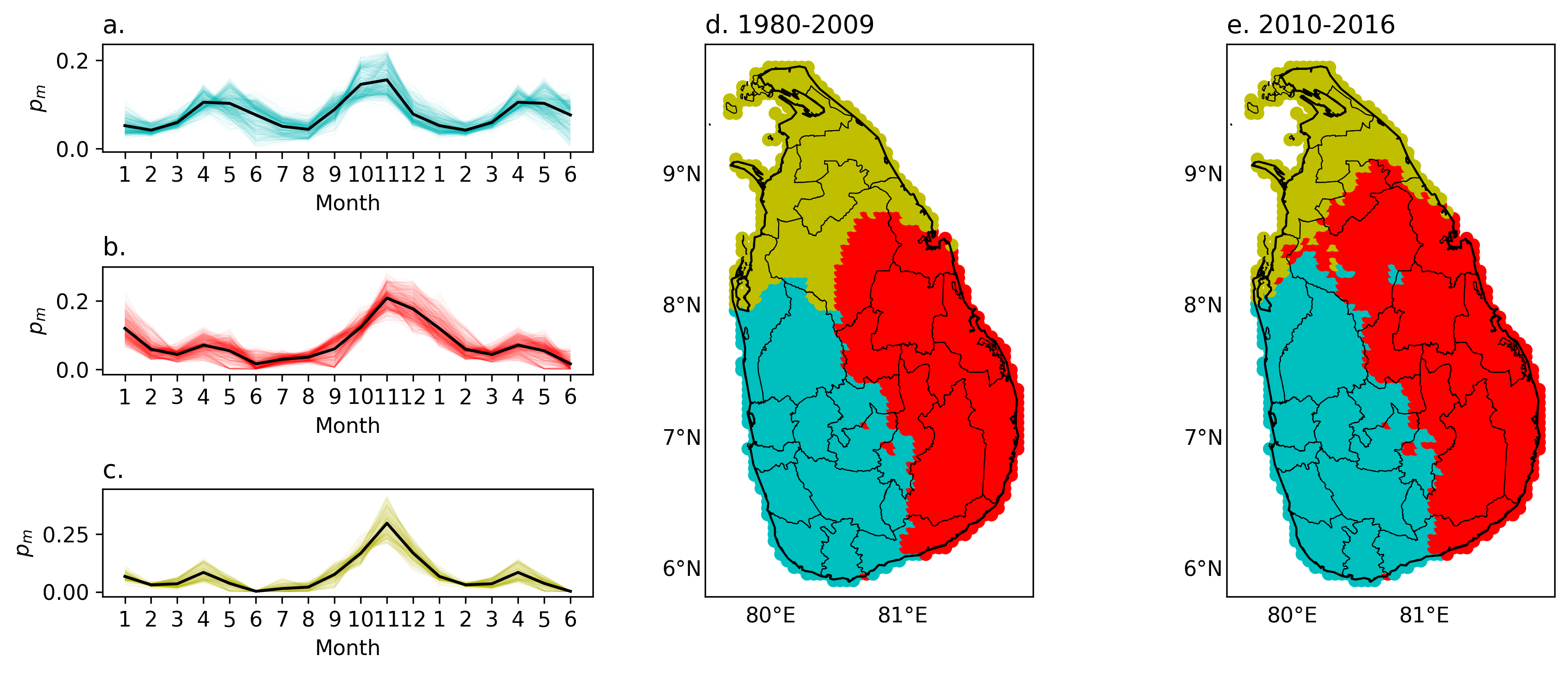}
    \caption{The seasonality of rainfall in Sri Lanka. (a-c) show three seasonality patterns extracted by clustering the vector $p$ where $p_m$ is the ratio of total rain in month $m$ to the total rain during 1980-2009. The centroid of seasonality patterns are shown as black solid lines. (d) and (e) are the spatial extent of each seasonality patterns during 1980-2009 and 2010-2016, respectively. The color codes corresponds to the pattern in (a-c).}
    \label{fig:Season}
\end{figure}

K-means is a clustering method that categorizes n-dimensional data points (here $n=12$) into $k$ clusters (here $k=3$) by minimizing the sum of the variance within clusters. Fig \ref{fig:Season}a-c show three distinct RSPs across Sri Lanka from 1980 to 2009. Each seasonality pattern is characterized by a mean (centroid of the cluster defined as the average of all cluster members) shown as solid black lines. The RSP at each location is then assigned to the cluster with a minimum Euclidean distance from its centroid. The spatial extent of these RSPs is shown in Fig. \ref{fig:Season}d. 

The rainfall in the northern region ( Fig. \ref{fig:Season}c)  exhibits an (almost) 'unimodal' seasonality; this is largely driven by the dominance of SIM spanning from October through December. The Eastern part of the island exhibits a fairly different seasonality pattern with an expanded winter rainfall from SIM  (from October to December) and NEM (from December to February), as shown in  Fig. \ref{fig:Season}b. This region also has relatively little rainfall from the summer monsoon. The rainfall in the southwest peaks during October and December mainly due to the SIM, with a second peak expanding through March to June from the FIM and SWM (Fig. \ref{fig:Season}a).

We further examine the change in seasonality patterns after 2010. This is achieved  by comparing seasonality in 2010-2016 at each point with the centroid of the clusters (black lines in Fig. \ref{fig:Season}a-c) and assigning them to the one with the smallest Euclidean distance. As shown in Fig. \ref{fig:Season}e, the patterns exhibit changes after 2010, with the encroachment of the eastern and western patterns toward the north. This is in line with the analysis of inter-annual rainfall trend which revealed lower rainfall from FIM and NEM after 2010 which implies a greater relative contribution from the summer monsoon and thus a more pronounced bi-modal pattern. 

\begin{figure}[h]
    \centering
    \includegraphics[width=0.5\textwidth]{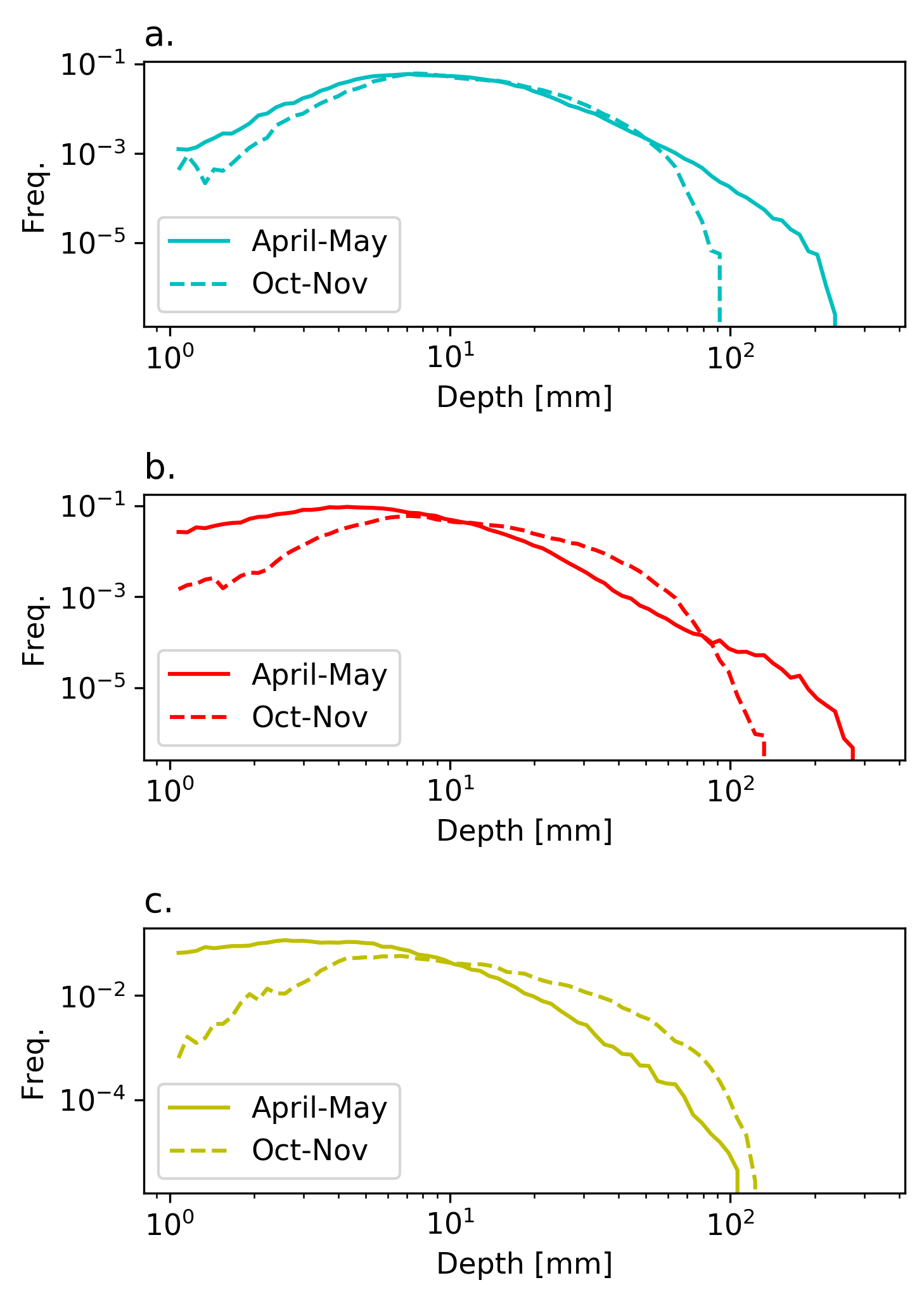}
    \caption{The probability distribution of the depth of rainfall events during two main seasons (April-May and October-November) in the areas associated with each RSP in Fig. \ref{fig:Season}a-c. The PDFs for events in April-May have an extended tails in the southern and eastern regions which implies more intense rain events.}
    \label{fig:depth}
\end{figure}

Rainfall seasons also exhibit diverse intermittency patterns. Fig \ref{fig:depth} shows the probability distribution of the depth of rain events during two main seasons (April-May and October-November) in the areas associated with each RSP in Fig. \ref{fig:Season}a-c. The total rainfall in the April-May season is generally less than that of October-November, as shown in Fig. \ref{fig:Season}a-c; however, in the eastern and southern regions, the rain events in April-May are relatively more intense, as highlighted by the extended tail in Fig. \ref{fig:depth}a and b. 

\section{Spatio-temporal dynamics of dengue fever in Sri Lanka}

As shown in the last section, rainfall in Sri Lanka exhibits spatially variables patterns associated with the four active monsoon seasons. Here we discuss the implication of these rainfall seasonality patterns in terms of the burden of dengue fever in Sri Lanka. 

\begin{figure}[h]
    \centering
    \includegraphics[width=0.7\textwidth]{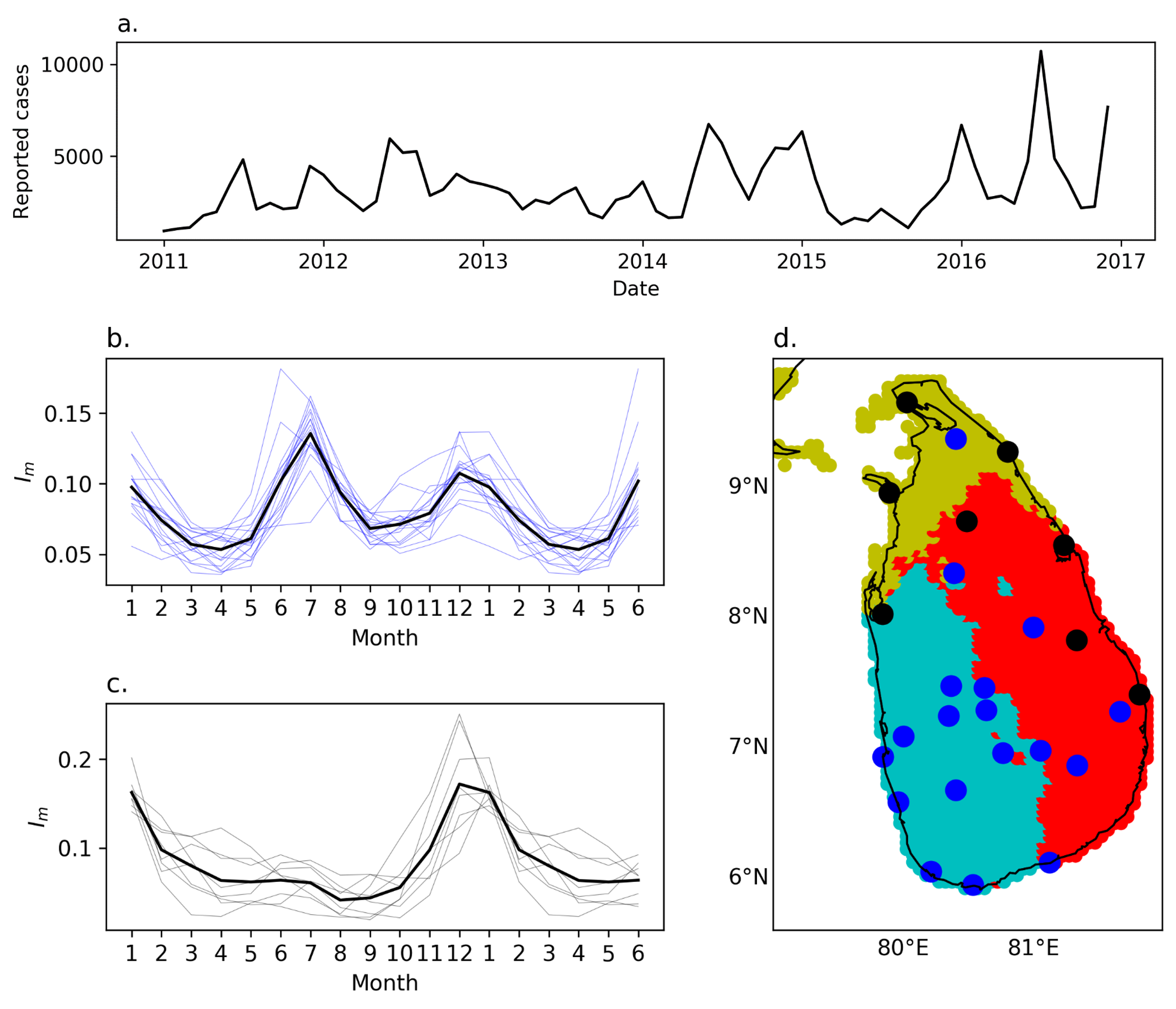}
    \caption{(a) Reported cases of dengue fever in Sri Lanka from 2010-2016. Data is from the Ministry of Health. (b) and (c) show two seasonality patterns extracted by clustering the vector $I$ where $I_m$ is the ratio of the number of cases in month $m$ to the total cases during 2010-2016. The centroid of seasonality patterns is shown as solid black lines. (d) shows the spatial extent of each seasonality pattern where the color codes correspond to the pattern in (b) and (c). The extent of rainfall seasonality pattern in the same period of time is shown in the background (same as Fig. \ref{fig:Season}e).}
    \label{fig:SeasonDeng}
\end{figure}

The co-circulation of all four serotypes of dengue fever is well-documented in the last 40 years \cite{kanakaratne2009severe}. Fig. \ref{fig:SeasonDeng} shows the number of monthly reported cases from 2010-2016 from Sri Lanka's Ministry of Health.  In 2017 there was a severe outbreak that has been associated with the emergence of a new strain \cite{wagner2020climatological, srilanla_action}, short-term migration, and changes in the breeding sites of the mosquitoes \cite{srilanla_action}. Thus, our analysis here is focused on the period 2010 to 2016 to avoid a bias due to the high number of cases in 2017.

 Fig. \ref{fig:SeasonDeng}a and b show two dengue seasonality patterns (DSP) that are extracted in a similar way as the rainfall patterns in Fig. \ref{fig:Season}. In this case, the clustering is performed on $I_m$ defined as the ratio of the number of cases in month $m$ to the total cases during 2010-2016. The corresponding locations of the patterns are marked at the coordinates of the districts capitals in Fig. \ref{fig:SeasonDeng}c. 

The DSPs exhibit uni- and bi-modal behaviors. The bi-modal pattern peaks in July and December and is prevalent mainly in south and southwest regions. The uni-modal pattern peaks in January and is spatially scattered throughout the island except for the southwestern part. The spatial extent of the rainfall seasonality pattern is shown in the background of Fig. \ref{fig:SeasonDeng}c. It is evident that the bi-modal DSP is mostly associated with the RSP in the southwest where two dominant rain season exists, although the dengue cases reach their maximum with an almost two-months delay relative to the rainfall peaks. The uni-modal DSP exists within regions with a relatively low contribution of spring rainfall in the eastern and northern regions. 

It is interesting to note that even without significant rainfall in April-May, dengue cases may peak during the summer in July. For instance, the two black dots in the northern regions in Fig. \ref{fig:SeasonDeng}b have a significant number of cases in July; however, their RSP is almost uni-modal with a relatively small rainfall during April-May. This suggests that there may be other hydro-climatic (humidity, temperature) or social (increase in human activities which can lead to more exposure to mosquito bites) factors that drive the dengue spread in that period. This is also reinforced by the fact that the smaller rainfall peaks in April-May (see Fig. \ref{fig:Season}a) leads to the larger dengue cases peaks in July, as shown in Fig. \ref{fig:Season}a.

\section{Modeling dengue dynamics with the HYSIR model}

We modeled the dengue transmission using a hydrologically driven SIR model \cite{hooshyar2020HYSIR}. In this formulation, the water availability $w$ is given by a simple bucket model given as  

\begin{equation}
    \label{eq:soil_mosisture}
   \frac{d w}{d t} = -\rho w + R(t), 
\end{equation}
where $w$ quantifies the water availability, $R\ [\text{d}^{-1}]$ is normalized (by a parameter $w_0$) rainfall rate. The parameter $\rho$ is proportional to $\frac{L_{max}}{w_o}$ where $L_max$ is the maximum rate of water loss and $\rho^{-1}$ captures the hydrologic 'memory' \cite{katul2007spectrum}. This type of bucket model has also been used to describe the temporal dynamics of soil moisture in the top soil layer \cite{porporato2004soil, rodriguez1999probabilistic}. In this study, we used $w_0=150\ \text{mm}$ although the choice of $w_0$ simply changes the scale of $w$ and does not alters the results presented here. We used $\rho=0.1\ \text{d}^{-1}$ which introduces a $10$-day memory in the $w$ dynamics.

Dengue transmission with a hydrologic component is modeled using the well-known SIR formulation \cite{keeling2011modeling,baker2019epidemic, grenfell2002dynamics, alonso2007stochastic},

\begin{equation}
    \label{eq:SIR}
    \begin{split}
        &\frac{dS}{dt} = -\mathcal{B}(w) I S - \eta S + \eta\\
        &\frac{dI}{dt} = \mathcal{B}(w)  I S - \eta I -\gamma I\\
        \end{split}
\end{equation}
where the ratio of susceptible, infected, and recovered individuals to the total population are denoted by $S$, $I$, and $R$, respectively. $\gamma$ is the recovery rate and $\eta$ is birth/death rate thus $\gamma^{-1}$ and $\eta^{-1}$ are the expected time of infection and individual life span, respectively.  The function $\mathcal{B}$  is the moisture-dependent transmission rate. 

\subsection{Reconstructing the transmission rate from incidence time-series}

\begin{figure}[h]
    \centering
    \includegraphics[width=1\textwidth]{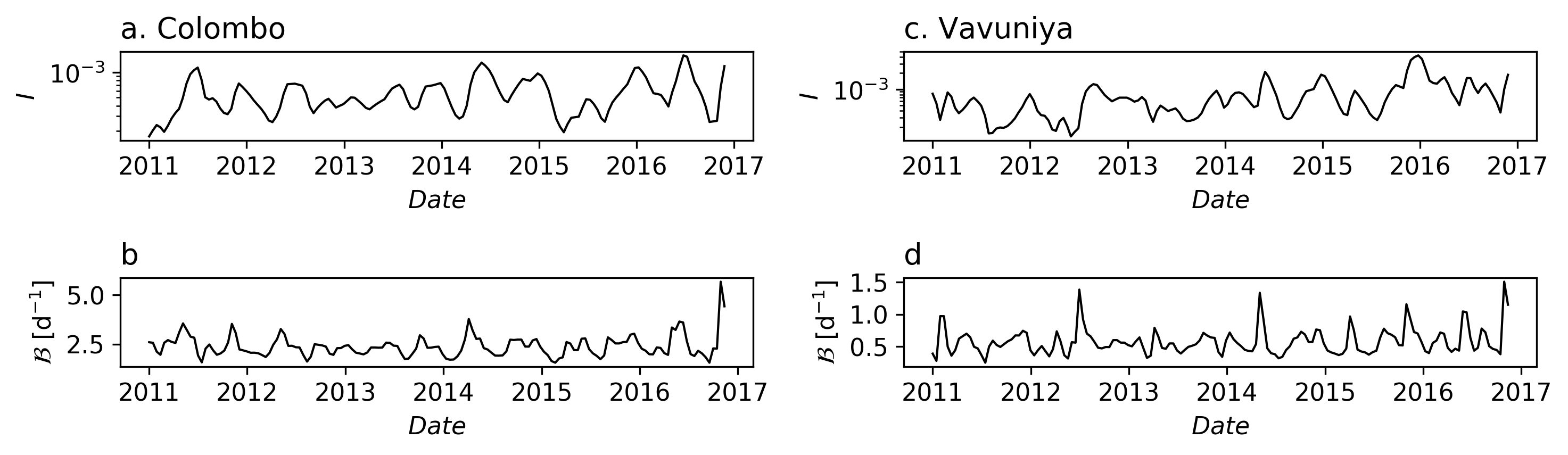}
    \caption{The reconstructed time-series of the incidence $I$ and transmission rate $\mathcal{B}$ in the districts Colombo and Vavuniya. These results are achieved from the TSIR method \cite{finkenstadt2000time, becker2017tsir} using the district-level reported cases, birth, and population time-series which were interpolated to bi-weekly temporal scale.}
    \label{fig:recons}
\end{figure}

The reported dengue cases in Sri Lanka are limited to incidence, whereas the number of susceptible individuals remains unobserved.  We used the TSIR method \cite{finkenstadt2000time, becker2017tsir} to reconstruct the 'true' times-series of cases and susceptible individuals. The TSIR uses the reported time-series of incidence, birth, and population to compute the reporting rate for the reconstruction of the 'true' incidence. The susceptible $S$ and transmission rate $\mathcal{B}$ time-series are then computed using a Generalized Linear Model (GLM) to find the best fit to the dynamics of $I$ given in Eq. \ref{eq:SIR}. The TSIR requires the frequency of recorded data to be equal to the recovery rate of disease which we take to be $1/14$ days as in \cite{wagner2020climatological}. Thus, we used simple linear interpolation to reconstruct bi-weekly time-series from monthly incidence, birth, and population data. Fig. \ref{fig:recons} shows the reconstructed incidence and transmission rate time-series in the districts Colombo and Vavuniya.    

\subsection{Linear model for transmission rate}

\begin{figure}[h]
    \centering
    \includegraphics[width=0.7\textwidth]{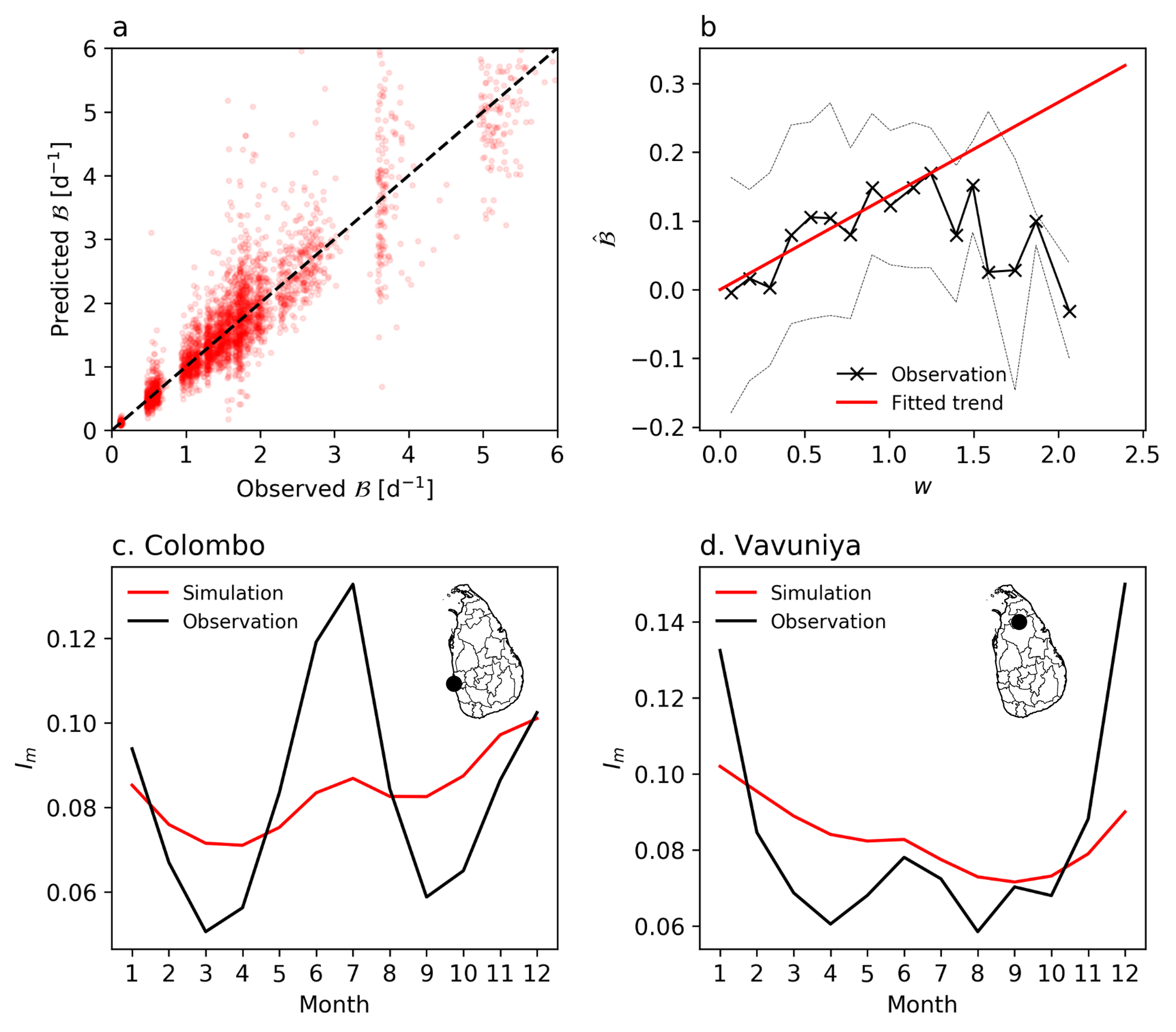}
    \caption{The performance of the linear model of transmission rate (see Eq. \ref{eq:beta_1}). (a) shows the observed versus predicted transmission rates as compared to the one-to-one trend expected for a perfect model. (b) shows the hydrologic controls on the transmission rate that is visualized by the relationship between $\mathcal{\hat{B}} = \ln \mathcal{B}-\zeta_d$ and $w$. The observations are binned and the mean, $25^{th}$, and $75^{th}$ envelops are shown. The fitted linear trend $\mathcal{\hat{B}} = \psi w$ is shown in red. (c) and (d) are seasonality of the incidence from observation and simulation in the districts Colombo and Vavuniya. The simulation are performed by forcing the HYSIR model with the observed rainfall and the transmission rate given in Eq. \ref{eq:beta_1}.}
    \label{fig:linearmodel}
\end{figure}

To study the impact of water availability on the transmission of dengue in Sri Lanka, we assume a relationship between $w$ and $\mathcal{B}$ of the form

\begin{equation}
    \label{eq:beta_1}
    \ln \mathcal{B} = \psi w + \zeta_d
\end{equation}
where the intercept $\zeta_d$ is defined for each district $d$. Eq. \ref{eq:beta_1} describes a linear relationship between the (log of) transmission rate and water availability given by water availability $w$. This simple linear dependence aims to capture the positive feedback of water availability on the mosquito life cycle, in terms of availability of breeding sites and improved survival. We should note that this formulation is a minimalist representation of a hydrologically driven mosquito-borne infection and neglects the explicit modeling of the mosquito life cycle \cite{rocha2013time}. The constants $\zeta_d$ quantify a non-seasonal control on transmission and capture the spatial variation of observed dengue cases across the island. Given the time-series of $\mathcal{B}$ (see Fig. \ref{fig:recons}b and d for examples), we used OLS \cite{hutcheson2011ordinary} to compute $\psi$ and $\zeta$ from Eq. \eqref{eq:beta_1}. Fig. \ref{fig:linearmodel}a shows the observed versus predicted transmission rate as compared to a one-to-one trend expected for the perfect model. 

The hydrologic impact on the transmission rate from Eq. \ref{eq:beta_1} can be better visualized by subtracting the spatial components $\zeta$ from the observed transmission rates ($\mathcal{\hat{B}} = \ln \mathcal{B}-\zeta_d$). The relationship between $\mathcal{\hat{B}}$ and $w$ is compared to the predicted linear trend (i.e., $\mathcal{\hat{B}} = \psi w$) in Fig. \ref{fig:linearmodel}b. For small $w$, the transmission rate linearly increases with $w$ in agreement with the functional form assumed in Eq. \ref{eq:beta_1}. However, we observed a deviation from this linear trend at high $w$. The reduction of transmission rate at high $w$ may be related to the adverse effects of flooding on mosquito life-cycle \cite{duchet2017effects, fouque2006aedes}, although other epidemiological feedbacks such as changes in human and mosquito activity patterns may play a role here. 

We also used Eq. \ref{eq:beta_1} with the fitted parameters to simulate the incidence from 2011 to 2016 and compared the seasonality patterns with the observation as shown in Fig. \ref{fig:linearmodel}c and d for the districts Colombo and Vavuniya. For these simulations, the initial conditions were set to the observed condition at $1/1/2011$. We used a simulated time-series of $w$ (using Eq. \ref{eq:soil_mosisture}) to compute $\mathcal{B}$ at each time step. Here, we performed a single deterministic simulation for each case and ignored the uncertainty in the fitted parameters by only considering the best-fitted values. Although the model captures the timing of incidence peaks accurately (see Fig. \ref{fig:linearmodel}c for instance), the relative magnitude of the peaks is poorly modeled. It should be noted that without any additional parameters, the 2-month phase difference between rainfall and dengue cases peaks is captured in this minimalist approach. This is due to the nonlinearity introduced by the multiplicative transmission term in Eq. \eqref{eq:SIR} as well as the hydrologic memory modeled by $\rho^{-1}$. 

\subsection{Quadratic model for transmission rate}

Inspired by the drop in transmission rate at high $w$ (Fig. \ref{fig:linearmodel}b), we explored a more flexible functional form for hydrologically driven transmission rate,

\begin{equation}
    \label{eq:beta_2}
    \ln \mathcal{B} = \psi w  + \theta w^2 + \zeta_d. 
\end{equation}

\begin{figure}[h]
    \centering
    \includegraphics[width=0.7\textwidth]{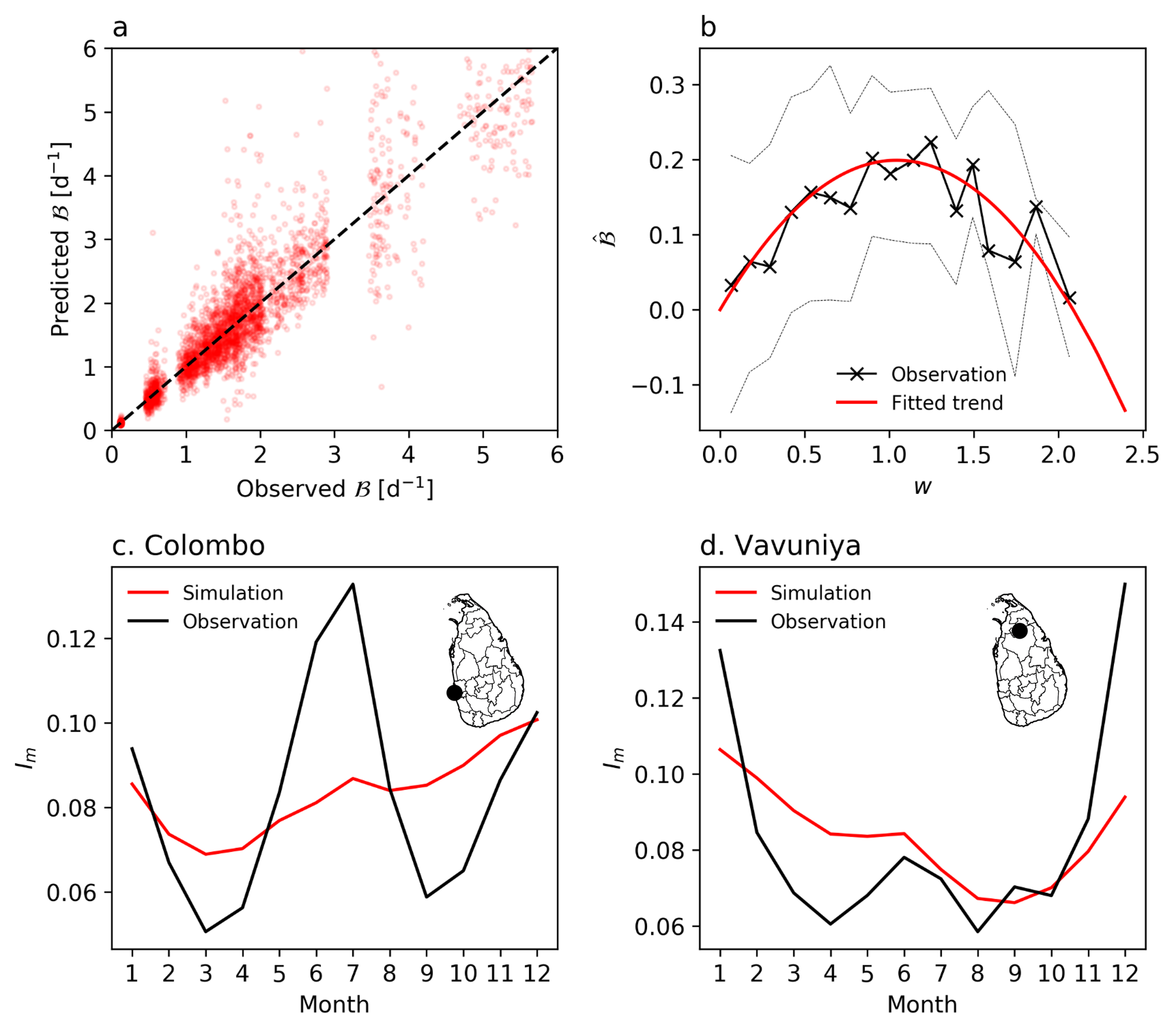}
    \caption{Same as Fig. \ref{fig:linearmodel} for the transmission model given in Eq. \ref{eq:beta_2}. }
    \label{fig:quadmodel}
\end{figure}

The quadratic term with coefficient $\theta$ allows for a non-monotonic dependence of the transmission rate on water availability. Fig. \ref{fig:quadmodel} shows the performance of the transmission model in Eq. \ref{eq:beta_2}. Although the hydrologic impact on the transmission rate is now captured more accurately, as compared to the linear model (Fig. \ref{fig:quadmodel}b), the magnitude of the seasonal peaks of incidence are still not fully resolved.

\subsection{Transmission model with seasonal coefficients}

\begin{figure}[h]
    \centering
    \includegraphics[width = 0.7\textwidth]{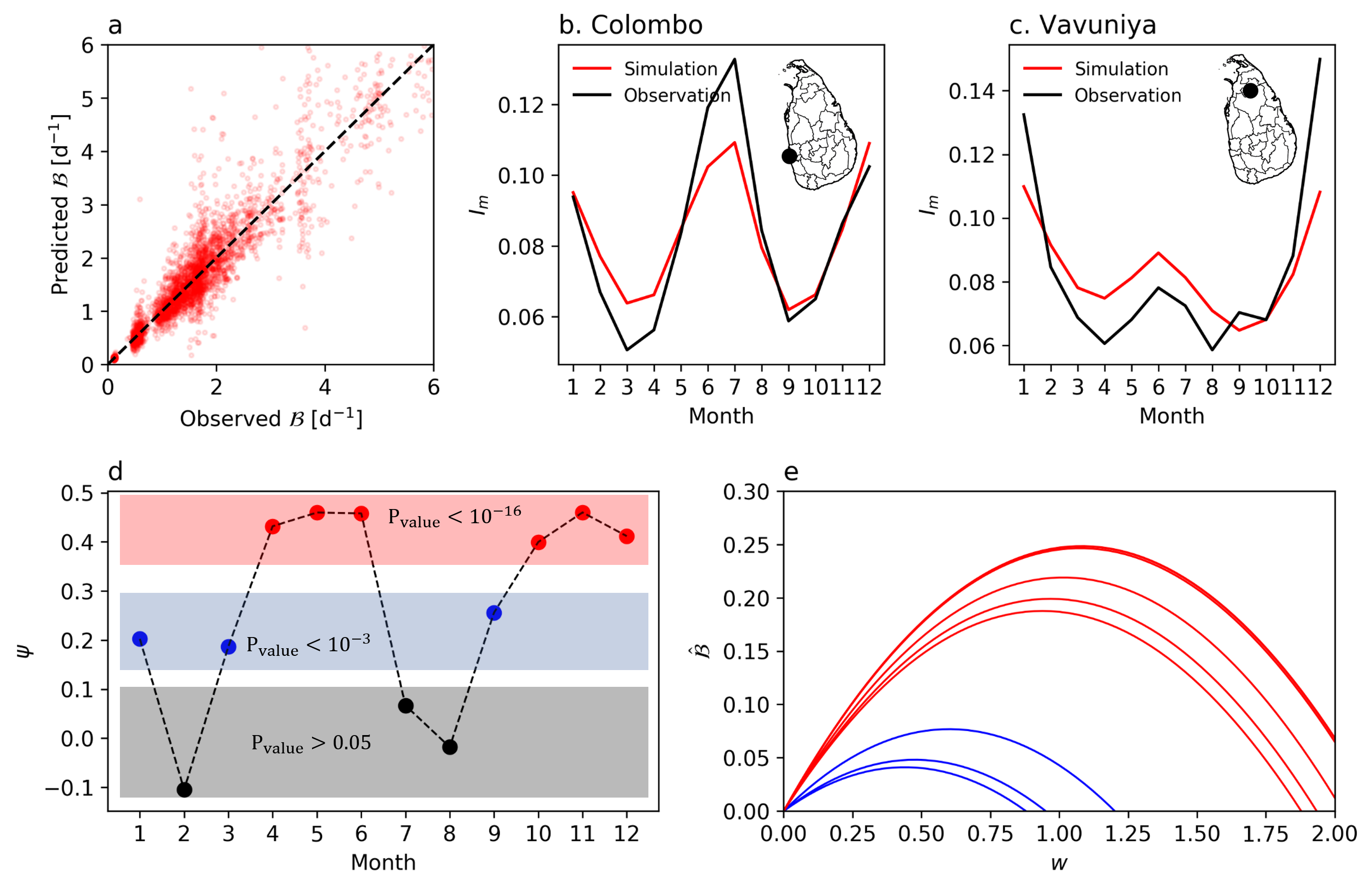}
    \caption{The performance of the model with the transmission rate given in Eq. \ref{eq:beta_monthly}: (a) shows observed versus predicted transmission rates as compared to the one-to-one trend expected for the perfect model;(b) and (c) are seasonality of the incidence from observation and simulation in the districts Colombo and Vavuniya (refer to Appendix I for the results of other districts. The simulation are performed by forcing the HYSIR model with observed rainfall and the transmission rate given in Eq. \ref{eq:beta_monthly}); (d) shows the monthly variation of the parameter $\psi$. The $\psi$ at February and August are not statistically significant ($p_{\text{Value}} >0.05$). The hydrologic controls on the transmission rate given by $\mathcal{\hat{B}} = \psi_m w + \theta w^2$ for each month is shown in (e). The colors corresponds to those shown in (d).}
    \label{fig:genmodel}
\end{figure}

To further improve the model performance, we modified the hydrologic portion of the transmission rate expression by allowing seasonal variation in the coefficient $\psi$,  

\begin{equation}
    \label{eq:beta_monthly}
    \ln \mathcal{B} = \psi_m w + \theta w^2 + \zeta_d,
\end{equation}
where $m$ is the index for the month. In this model, the coefficients $\psi$  are month-specific, the intercepts $\zeta$ are location-specific, and $\theta$ is a constant. Allowing for seasonal variation in $\psi$ enables a more flexible dependence of transmission rate on water availability at the monthly time scale. Fig. \ref{fig:genmodel}a shows the relationship between the observed and predicted transmission rate by Eq. \ref{eq:beta_monthly} after fitting to the observed time-series of $\mathcal{B}$. The fitted function is then used to simulate the incidence using the HYSIR model. The seasonality of observed versus simulated incidence in the districts Colombo and Vavuniya are shown in Fig. \ref{fig:genmodel}b and c. It is evident that the addition of the seasonal coefficients $\psi$ allows for a more accurate prediction of the seasonal peaks of dengue incidence. A similar comparison for all 21 studied districts is shown in Appendix I (Fig. \ref{fig:alldost}). 

The monthly variation of the parameter $\psi$ is shown in Fig. \ref{fig:genmodel}d which exhibits an almost symmetric bimodal pattern with peaks during April-July and October-December. This indicates a relatively more pronounced response of the transmission rate on water availability in those periods as also depicted by the red lines in Fig. \ref{fig:genmodel}e . On the other hand, the values of $\psi$ in February, July, and August are not statistically significant ($p_{\text{Value}}>0.05$) which indicates that the transmission rate is unresponsive to water availability in those periods. At the transition between these two regimes, the transmission rate moderately increases with water availability before dropping at high values of $w$ (the blue lines in Fig. \ref{fig:genmodel}e). 

\section{Conclusion}

Understanding the role of hydro-climatic conditions on the spread of vector-borne infections is essential for quantifying the risk of these health hazards, especially in the face of the future changes in patterns \cite{wagner2020climatological}. With this in mind, we studied dengue transmission in Sri Lanka and showed that the hydrology plays a major role in the observed seasonality pattern of dengue incidence. This analysis highlights the importance of the seasonality of climatic variables (e.g., rainfall) which are important drivers of infection dynamics. These seasonal patterns have been have been changing rapidly  \cite{feng2013changes} and are expected to continue doing so, especially in tropics where the burden of vector-borne infections is the highest \cite{bhatt2013global} . 

Our analysis also shows a non-monotonic dependence of dengue transmission rate on water availability $w$. The initial increase in the transmission rate with high $w$ may be related to the positive feedback of water availability on the mosquito life cycle, which leads to an abundance of breeding sites and a better chance of survival. A similar increasing trend has been observed in mosquito populations \cite{carvajal2018machine, hii2012forecast}; however, the opposite trend has also be reported in the analysis of dengue cases with river level in Bangladesh \cite{hashizume2012hydroclimatological}. The effect of intense floods and flash-floods can be more complicated and difficult to predict. Flash-floods may disturb the mosquito life-cycle, destroy their habitats, and reduce their food resources which may lead to a decrease in their population size \cite{duchet2017effects, fouque2006aedes}. This can explain the decrease in transmission rate at high $w$ observed in this study. For instance, Lehman et al. \cite{lehman2007effect} argued that the heavy flooding and strong winds of Hurricane Katrina might have actually decreased the risk of mosquito-borne diseases by dispersing and destroying mosquito habitat in Louisiana and Mississippi, USA. However, at longer time-scales, flash-floods may have positive feedback effects on the abundance of mosquitoes as they can reduce the population of mosquito antagonists \cite{duchet2017effects}. 

Although the rainfall seasonality was shown to drive the seasonality of dengue cases in Sri Lanka, even in the absence of climatic seasonality, the HYSIR model exhibits noise-induced cyclic behavior that is controlled by the hydrologic memory introduced by parameter $\rho$ \cite{hooshyar2020HYSIR}. Here we assumed a constant $\rho$; however, accounting for the spatial and temporal variation of $\rho$ can potentially improve the model predictions. Our analysis focused on water availability and rainfall, although other hydro-climatic variables such as temperature, humidity, wind speed, etc. may be important as well \cite{gharbi2011time, lu2009time, wagner2020climatological, mordecai2017detecting, mordecai2019thermal}.  

Our model for transmission rate assumes a dominant linear dependence on water availability with slope $\psi$ at low values of $w$. This parameter varies seasonally in order to capture the seasonal peak of dengue cases. The parameters $\psi$ is a surrogate for a range of controlling factors such as the abundance of 'potential' breeding sites (e.g., the density of buckets regardless of their water content). We should also note that the value of $\psi$ may as well reflect monthly variations in other aspects of the disease spread, such as changes in human-mosquito activity and contact processes, etc. For instance, the observation of two annual peaks in $\psi$ coincides with the sowing and growing periods of Yala and Maha cultivation seasons in Sri Lanka \cite{fao}.

\section{Acknowledgements}

We acknowledges support from Princeton Environmental Institute and Princeton Institute for International and Regional Studies at Princeton University though the climate and disease initiative. The numerical simulations in this article were performed on computational resources provided by Princeton Research Computing, a consortium of groups including the Princeton Institute for Computational Science and Engineering (PICSciE) and the Office of Information Technology's High Performance Computing Center and Visualization Laboratory at Princeton University.

\section{Data}
We used precipitation data from the Climate Hazards group Infrared Precipitation with Stations (CHIRPS) with a spatial resolution of 0.05$^{\circ}$ at a daily time scale \cite{funk2015climate}. The monthly reported case of dengue in Sri Lanka is from the official website of the epidemiology unit of the Ministry of Health of Sri Lanka \cite{epid}. We birth and population data were acquired from the Department of Census and Statistics of Sri Lanka \cite{srilankacencus}.

\section{Appendix I}

\begin{figure}[H]
    \centering
    \includegraphics[width=0.9\textwidth]{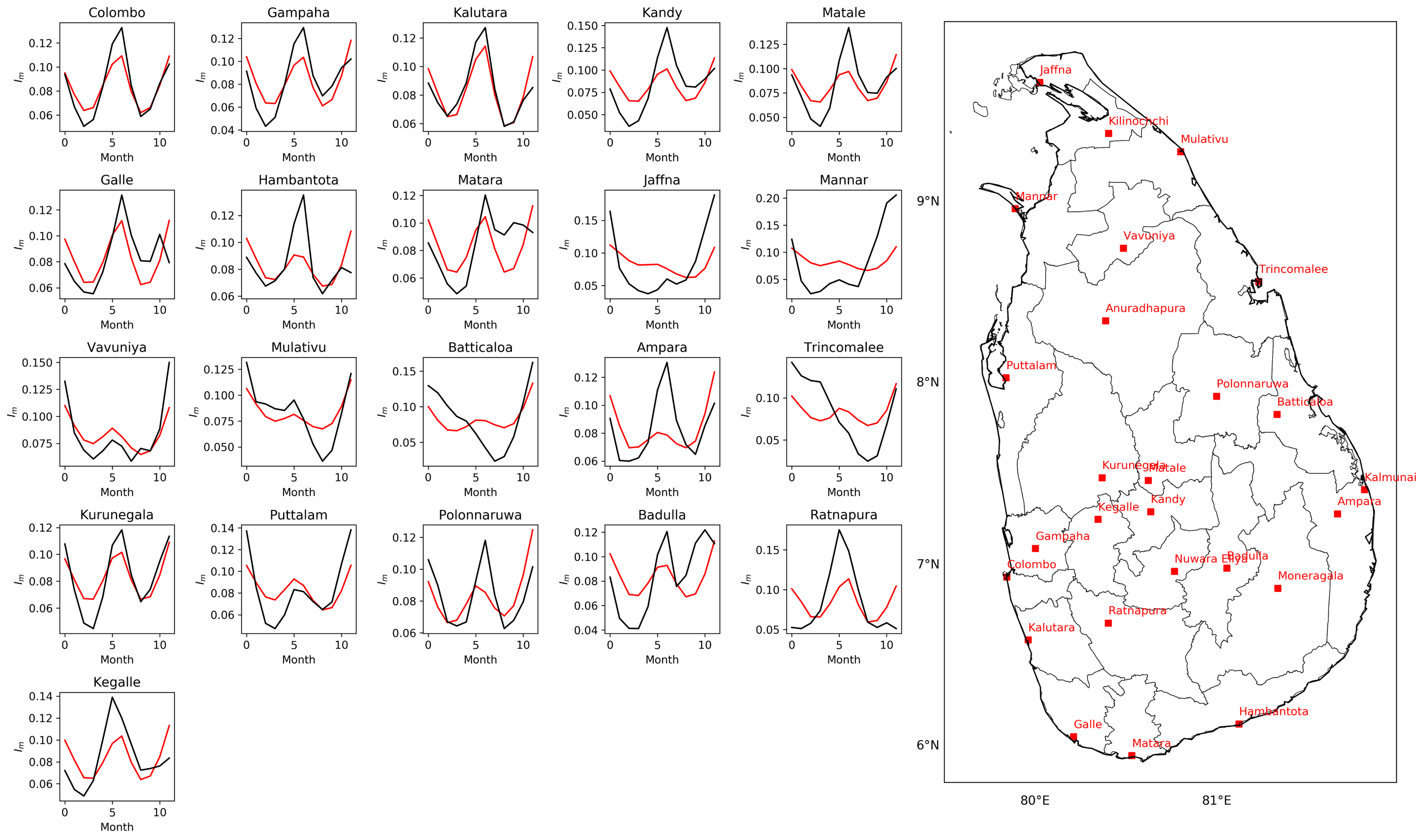}
    \caption{The comparison of incidence seasonality from observation (black lines) of simulation (red lines) in the studied districts. The transmission rate is given by Eq. \eqref{eq:beta_monthly}. }
    \label{fig:alldost}
\end{figure}

The observed and simulated seasonality patterns of dengue incidence from the quadratic model in Eq. \eqref{eq:beta_monthly} is shown in Fig. \ref{fig:alldost}. 

\bibliographystyle{unsrt}
\bibliography{ref.bib}

\end{document}